\documentclass[10pt, draftclsnofoot, peerreview, letterpaper, onecolumn]{IEEEtran}
\usepackage{cite}
\usepackage{graphicx}
\usepackage{amsmath}
\usepackage{amsfonts}
\title{How to Fully Exploit the Degrees of Freedom in the Downlink of MISO Systems With Opportunistic Beamforming?}
\author{Minghua~Xia, Wenkun~Wen, and Soo-Chang Kim
\thanks{This work was supported in part by the MKE/IITA of Korea under the IT R\&D program 2006-S001-02, Development of Adaptive Radio
Access and Transmission Technologies for 4th Generation Mobile
Communications.}
\thanks{Minghua Xia and Soo-Chang Kim are with the ETRI Beijing
R\&D Center, Beijing 100027, China (e-mail:
xia\_minghua@yahoo.com.cn; sckim@etri.re.kr).}
\thanks{Wenkun~Wen is with Guangdong-Nortel R\&D Center, Guangzhou 510665, China (e-mail:
wenkun@gmail.com).}}

\begin{document}

\maketitle

\begin{abstract}
The opportunistic beamforming in the downlink of multiple-input
single-output (MISO) systems forms $N$ transmit beams, usually, no
more than the number of transmit antennas $N_t$. However, the
degrees of freedom in this downlink is as large as $N_t^2$. That is,
at most $N_t^2$ rather than only $N_t$ users can be simultaneously
transmitted and thus the scheduling latency can be significantly
reduced. In this paper, we focus on the opportunistic beamforming
schemes with $N_t<N\le N_t^2$ transmit beams in the downlink of MISO
systems over Rayleigh fading channels. We first show how to design
the beamforming matrices with maximum number of transmit beams as
well as least correlation between any pair of them as possible,
through Fourier, Grassmannian, and mutually unbiased bases (MUB)
based constructions in practice. Then, we analyze their system
throughput by exploiting the asymptotic theory of extreme order
statistics. Finally, our simulation results show the
Grassmannian-based beamforming achieves the maximum throughput in
all cases with $N_t=2$, $3$, $4$. However, if we want to exploit
overall $N_t^2$ degrees of freedom, we shall resort to the Fourier
and MUB-based constructions in the cases with $N_t=3$, $4$,
respectively.
\end{abstract}


\newpage

\section{Introduction} \label{Section 1}
\PARstart{M}{ultiple-input} multiple-output (MIMO) system holds
promise for the next generation wireless communications due to its
high spectral efficiency \cite{Gesbert03,Paulraj04}. In a
single-user MIMO system, its capacity has been extensively
investigated, assuming different channel state information (CSI) is
known at the transmitter and/or receiver
\cite{Foschini98,Telatar99}. In the multi-user scenario, the
multi-user diversity was introduced as a new dimension of degrees of
freedom to further increase the capacity
\cite{Knopp95,Tse97,Bender00,Borst01}. In this paper, we focus on
the downlink of multi-user MIMO systems, i.e., broadcast channels
(BCs). By using dirty paper coding strategy at the transmitter
\cite{Costa83}, the optimal sum-rate capacity region of MIMO BCs was
well established from the information-theoretic viewpoint, with the
assumption that CSI is perfectly known at the transmitter and all
the receivers \cite{Caire03,Viswanath03-1}. This region can be
numerically evaluated by using the duality between BCs and
multi-access channels (MACs), though it is extremely computationally
intensive \cite{Jindal04,Vishwanath03-2,Jindal05}. In practice, the
optimal sum-rate capacity region of MIMO BCs can be approached using
the nested lattices or trellis beamforming scheme
\cite{Zamir02,Yu05}, which generalizes the Tomlinson-Harashima
beamforming \cite{Tomlinson71, Harashima72}. Unfortunately, perfect
CSI at the transmitter is almost infeasible in practical
communication systems with large number of users, and also the
non-linear beamforming is usually impractical for the real-time
traffic. Therefore, designing linear beamforming schemes with lower
feedback complexity is of great interest
\cite{Gesbert04,Gesbert07,Vu07}.

The opportunistic beamforming system (OBS), also known as the random
beamforming system, is shown in \cite{Bayesteh07} to achieve the
maximum sum-rate capacity with the minimum amount of feedback,
provided that the number of users is not smaller than the number of
transmit antennas. This condition is surely satisfied in the
practical cellular systems. The single-beam OBS is proposed in
\cite{Viswanath02}, in which the conceptual idea of multi-beam OBS
is also presented in \cite[Appendix B]{Viswanath02}. The detailed
analysis on the throughput\footnote{In this paper, the term
``throughput'' refers to the average sum rate capacity, and the
link-adaptive techniques, such as adaptive coding/modulation, finite
constellation and dynamic power allocation, are not taken into
account.} of OBS with multiple orthogonal transmit beams is
performed in \cite{Sharif05,Sharif07}. Moreover, the opportunistic
beamforming with only signal-to-interference-plus-noise ratio (SINR)
feedback is generalized in \cite{Yoo07} to the case with composite
feedback consisting of quantized channel directional information and
channel quality information (channel magnitude or SINR).

In the literature with respect to multi-beam OBS
\cite{Sharif05,Sharif07,Yoo07,Jiang04,Ratnarajah07,Hochwald04},
random vector quantization (RVQ) limited feedback MIMO systems
\cite{Jindal06,Srinivasa07,Jafar07,Yoo07}, or the 3GPP Long Time
Evolution (LTE) of 3G systems \cite{3GPP08}, it is always assumed
that the number of transmit beams $N$ is identical to the number of
transmit antennas $N_t$ and thus there are at most $N_t$ users can
be simultaneously transmitted. In other words, the beamforming
matrix is a square matrix with size $N_t \times N_t$. However, it is
shown that the optimal transmission strategy regarding the
sum-capacity criterion in MIMO broadcast channels with large number
of users involves more than $N_t$ transmit beams at the same time
but upper bounded by $N_t^2$, i.e., $N_t<N \le N_t^2$ \cite{Yu06}.
If each user is equipped with $N_r>1$ receive antennas, he/she can
receive up to $N_r^2$ data streams \cite{Yu06}. This allows the user
to increase his/her own data rate but it prevents the simultaneous
transmission by other users, so that the number of simultaneously
transmitted users is limited to be $\lceil N_t^2/N_r^2\rceil$, where
we assume that each user receives exactly $N_r^2$ data streams and
$\lceil x\rceil$ denotes the integer ceiling operator \cite{Yu06}.
In other words, there are at most $N_t^2$ degrees of freedom in the
extreme case with $N_r=1$. Throughout this paper, we suppose there
is only $N_r=1$ receive antenna for each user and thus there are at
most $N_t^2$ users that can be simultaneously transmitted. The
beamforming matrix is now oblong with size $N_t \times N_t^2$.
Unfortunately, \cite{Yu06} does not show us the implementation of
beamforming schemes with $N_t<N \le N_t^2$ simultaneously
transmitted users. To the best of authors' knowledge, only the case
with $N=N_t+1$ scheduled users is addressed in \cite{Zorba08} by
exploiting the tight Grassmannian frames.

In this paper, we show how to schedule $N_t<N \le N_t^2$ users
simultaneously. Specifically, we design the opportunistic
beamforming schemes with $N_t<N \le N_t^2$ transmit beams in which
one user is scheduled at each beam. In particular, $N_t$ is supposed
no larger than $4$, just as that in 3GPP LTE \cite{3GPP08}. Unlike
the orthogonal transmitting case \cite{Sharif05}, the orthogonality
between different transmit beams is not retained again if $N>N_t$.
More precisely, the rank of beamforming matrix $\boldsymbol{B}\in
\mathbb{C}^{N_t \times N}$ is certainly no larger than $N_t$, where
$\mathbb{C}$ stands for the field of complex numbers. That is, there
are at least $N-N_t$ transmit beams that are no longer be orthogonal
with the others. However, if the transmit beams are generated as at
least correlated as possible, more transmit beams benefit to
schedule more users as soon as possible and hence decrease the
scheduling latency. Unfortunately, the increased multi-user
interferences and the loss of orthogonality between transmit beams
will inevitably deteriorate system throughput. Therefore, there must
be a tradeoff between more and more transmit beams and increased
multi-user interferences as well as disappearing orthogonality. In
this paper, we first show how to construct the beamforming matrices
and then the system throughput is rigorously investigated.

The rest of this paper is organized as follows. We present the
system model and scheduling strategy in Section \ref{Section 2}. In
Section \ref{Section 3}, we show how to design the beamforming
matrices with constrained correlation property. Then, the system
throughput is analyzed in Section \ref{Section 4}. Simulation
results and discussion are presented in Section \ref{Section 5}, and
finally, Section \ref{Section 6} concludes the paper.

\section{System Model And Scheduling Strategy}  \label{Section 2}
\subsection{System Model}
In this paper, we consider the downlink of a homogeneous single-cell
cellular system where the base station with $N_t$ antennas transmits
packets to $K$ single-antenna users, that is, the number of receive
antennas $N_r=1$ for each user. The number of users $K$ is assumed
no less than $N_t^2$ and all users are scattered geographically and
do not cooperate;\footnote{The case in which the number of users and
the number of transmit antennas are of the same order is addressed
in \cite{Jindal06,Dai08} and the references therein.} moreover,
their average SNRs are identical. Block flat Rayleigh fading
channels are supposed and all the time indices are omitted for the
sake of notation brevity if no other specific statement;
furthermore, different channels among users are mutually
independent. In addition, we suppose that the transmission time is
divided into consecutive and equal time slots, and each time slot is
less than the possible time delay but long enough so that there is a
coding strategy available that operates closely to Shannon channel
capacity. Moreover, each time slot is divided into a number of equal
sized mini-slots and several initial ones are used to transmit
common pilot symbols, so that the base station can determine which
users shall be chosen for data transmission in the rest mini-slots
according to the feedback of each user.

\begin{figure}
\centering
\includegraphics [width=4in,clip,keepaspectratio]{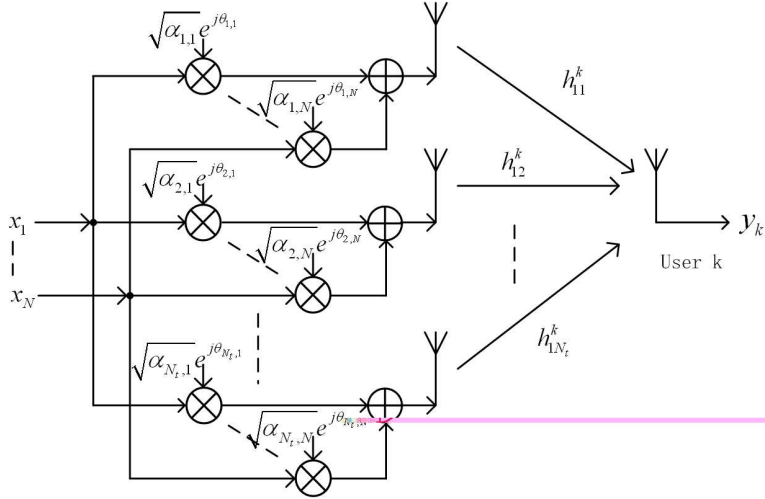}
\caption{The block diagram of OBS with multiple transmit beams}
\label{Fig.1}
\end{figure}

The OBS with $N$ transmit beams is illustrated in Fig.~\ref{Fig.1}.
At the base station, $N$ different beams are simultaneously
transmitted during one time slot, where $N\in [1,\,N_t^2]$. When
$N=1$, it denotes the single-beam transmission \cite{Viswanath02}.
When $N=N_t$, it refers to the conventional multi-beam orthogonal
transmission\cite{Sharif05, Sharif07, Yoo07}. In this paper, we
concentrate on the cases with $N_t< N \le N_t^2$.

When pilot symbols are transmitted during the first several
mini-slots,\footnote{It is shown that on an average $2.5$ mini-slots
is required to find the scheduled users \cite{Qin06}.}
$\boldsymbol{x}=[x_{_1},\cdots,x_{_N}]^T\in \mathbb{C}^{N\times 1}$
comprises $N$ different elements simultaneously transmitted at $N$
different beams. Furthermore, $x_{_n},\,n=1,\cdots,N$ is
simultaneously sent out from $N_t$ transmit antennas with each being
multiplied by a beamforming coefficient
$\sqrt{\alpha_{_{i,n}}}e^{j\theta_{i,n}}$ at Antenna $i,\,1\le i\le
N_t$ and Beam $n,\,1\le n\le N$. That is, $N$ different pilot
symbols are needed to distinguish $N$ transmit beams. On the other
hand, when data is transmitted during the later mini-slots, $x_{_n}$
refers to user data transmitted at Beam $n$. Then, the received
symbol of User $k$, $y_{_k}\in \mathbb{C}^{1\times 1}$, is given by

\begin{equation}\label{Eq.2-1}
y_{_k}=\sqrt{\frac{\rho}{N}}\sum_{n=1}^{N}{\boldsymbol{h}_{_k}\boldsymbol{b}_{_n}x_{_n}}+z_{_k}
\end{equation}
where $\rho$ is the average received SNR for each user;\footnote{In
order to make a fair comparison between different transmit schemes,
the transmit power in our proposal is normalized such that it is
independent of the number of transmit beams $N$ as shown in
(\ref{Eq.2-1}), whereas in the conventional orthogonal opportunistic
beamforming systems the transmit power is assumed to be identical
with $N$ \cite[Footnote 3]{Sharif05}.} $z_{_k} \in \mathbb{C}^{1
\times 1}$ stands for additive white Gaussian noise with zero mean
and unit variance; $\boldsymbol{h}_{_k} \in \mathbb{C}^{1 \times
N_t}$ denotes the complex channel vector between User $k$ and the
base station, and it agrees with Rayleigh fading with zero mean and
variance $1/m$. Moreover, the instantaneous beamforming matrix
$\boldsymbol{B}(t)\in \mathbb{C}^{N_t\times N}$ at time slot $t$ can
be written as
\begin{equation}\label{Eq.2-3}
\boldsymbol{B}(t)=\left[
                   \begin{array}{cccc}
                     \boldsymbol{b}_{_1} & \boldsymbol{b}_{_2} & \cdots & \boldsymbol{b}_{_N} \\
                   \end{array}
                 \right]
\end{equation}
where
\begin{equation}\label{Eq.2-4}
\boldsymbol{b}_{_n}=\left[
                   \begin{array}{cccc}
                     \sqrt{\alpha_{_{1,n}}}e^{j\theta_{1,n}} & \sqrt{\alpha_{_{2,n}}}e^{j\theta_{2,n}}
                      & \cdots & \sqrt{\alpha_{_{N_t,n}}}e^{j\theta_{N_t,n}} \\
                   \end{array}
                 \right]^T
                 ,\quad n=1,\,\cdots,\,N
\end{equation}
is the beamforming vector at Beam $n$, and $(.)^T$ denotes the
transpose operator.

In the case with $N=N_t$, the amplitudes
$\alpha_{_{i,n}},\,i=1,\cdots,N_t$ in (\ref{Eq.2-4}) are uniformly
distributed over $[0,1)$ such that
$\sum_{i=1}^{N_t}\alpha_{_{i,n}}=1$, and the phases
$\theta_{_{i,n}},\,i=1,\cdots,N_t$ are independent and uniformly
distributed over $[0,2\pi)$. Moreover, different beamforming vectors
are orthogonal with each other, that is,
\begin{equation}\label{Eq.2-5}
\boldsymbol{b}_{_l}^H\boldsymbol{b}_{_n}=\left\{
                                     \begin{array}{ll}
                                       1, & \hbox{$l=n$} \\
                                       0, & \hbox{$l\neq n$}
                                     \end{array}
                                   \right.
\quad l,\,n=1,\,\cdots,\,N_t
\end{equation}
where $(.)^H$ denotes the Hermitian transpose operator. Now,
$\boldsymbol{B}(t)$ is a unitary matrix and it can be generated
according to an isotropic distribution. However, in the cases with
$N_t<N\le N_t^2$, the beamforming vectors
$\boldsymbol{b}_{_n},\,n=1,\cdots,N$ are no longer orthogonal with
each other.

\subsection{Scheduling Strategy}
As far as the scheduling strategy is concerned, we assume that each
user sends them back to the base station, his/her maximum received
SINR and its corresponding beam index among $N$ different
beams.\footnote{Actually, it is not necessary for each user to offer
his/her feedback information. Instead, the same system throughput
can be nearly achieved by only allowing the strongest $10\%$ users
to provide feedback, whose received SINRs are above a predefined
threshold level. This is the so-called selective multi-user
diversity beneficial to greatly decrease the feedback complexity
\cite{Gesbert04}.} The feedback beam-index $\hat{n}_{_k}\in [1,\,N]$
of User $k$ is determined by
\begin{equation}\label{Eq.2-6}
\hat{n}_{_k}=\arg\max\limits_{n=1,\,\cdots,\,N}|\boldsymbol{h}_{_k}\boldsymbol{b}_{_n}|
\end{equation}
where $|x|$ denotes the amplitude of $x$.

According to (\ref{Eq.2-1}), the received SINR of User $k$ at Beam
$n$ is
\begin{equation}\label{Eq.2-7}
\gamma_{_{n,\,k}}=\frac{\frac{\rho}{N}|\boldsymbol{h}_{_k}\boldsymbol{b}_{_n}|^2}{1+\frac{\rho}{N}\sum\limits_{l=1,\,l\neq
n}^{N}{|\boldsymbol{h}_{_k}\boldsymbol{b}_{_l}|^2}}
\end{equation}
Hence, his/her maximum SINR among $N$ beams is
\begin{equation} \label{Eq.2-8}
\hat{\gamma}_{_k}=\max\limits_{n=1,\,\cdots,\,N}\gamma_{_{n,\,k}}
\end{equation}

Combining (\ref{Eq.2-6}) and (\ref{Eq.2-8}), the feedback
information of User $k$ can be shown as
$\left(\hat{n}_{_k},\hat{\gamma}_{_k}\right)$. At the base station,
there are $N$ different data sets $\mathcal{S}_n$, where $n \in
[1,\,N]$, corresponding to $N$ different beams to store the feedback
information of all users. That is, for any feedback
$\left(\hat{n}_{_k},\hat{\gamma}_{_k}\right)$, if $\hat{n}_{_k}=n$,
then $\hat{\gamma}_{_k} \in \mathcal{S}_n$. Moreover, the maximum
SINR scheduling strategy is adopted at the base station to choose a
user for transmission at each beam. Therefore, the index of
scheduled user at Beam $n$ is
\begin{equation}\label{Eq.2-9}
\hat{k}_{_n}=\arg\max\limits_{\hat{\gamma}_{_{k}} \in \mathcal{S}_n
}\hat{\gamma}_{_{k}},\quad n=1,\,\cdots,\,N
\end{equation}
Finally, the maximum SINR directing the transmission at Beam $n$ is
$\hat{\gamma}_{_{\hat{k}_{_n}}}$.

\section{The Implementation of OBS With $N_t<N \le N_t^2$} \label{Section 3}
In this section, we show three practical beamforming schemes with
$N_t<N \le N_t^2$. In general, the instantaneous beamforming matrix
$\boldsymbol{B}(t)$ shown in (\ref{Eq.2-3}) is constructed with a
fixed initial matrix $\boldsymbol{B}\in \mathbb{C}^{N_t\times N}$
and a time-variable vector
\begin{equation} \label{Eq.3-1}
\boldsymbol{c}(t)=\begin{bmatrix}
      e^{j\theta_1} & e^{j\theta_2} & \cdots & e^{j\theta_N} \\
    \end{bmatrix}
\end{equation}
in which $\theta_n,\,n=1,\cdots,N$ are fixed in time slot $t$ but
varied from time slot $t$ to $t+1$, furthermore, they are
independent and uniformly distributed over $[0,\,2\pi)$. More
accurately, $\boldsymbol{B}(t)$ is generated as
\begin{eqnarray}
\boldsymbol{B}(t)   &   =  &  \begin{bmatrix} \label{Eq.3-2}
      \boldsymbol{b}_{_1} & \boldsymbol{b}_{_2} & \cdots &   \boldsymbol{b}_{_N}
    \end{bmatrix} \\
         &   =   & \begin{bmatrix} \label{Eq.3-3}
      e^{j\theta_1}\boldsymbol{B}(:\,,\,1) & e^{j\theta_2}\boldsymbol{B}(:\,,\,2) & \cdots &
      e^{j\theta_N}\boldsymbol{B}(:\,,\,N)
    \end{bmatrix}
\end{eqnarray}
where $\boldsymbol{B}(:\,,\,n)$ refers to the $n^{th}$ column of
$\boldsymbol{B}$. Equation (\ref{Eq.3-3}) implies only a phase
rotation is performed on each column of $\boldsymbol{B}$ to get
$\boldsymbol{B}(t)$. Therefore, the correlation property of
$\boldsymbol{B}$ is remained.

From a purely information-theoretic point of view, using the
deterministic initial beamforming matrix $\boldsymbol{B}$ yields the
same system throughput as that if the time-variable
$\boldsymbol{B}(t)$ shown in (\ref{Eq.3-3}) is applied in fast
fading environment. The artificial randomness introduced by
$\boldsymbol{c}(t)$ shown in (\ref{Eq.3-1}) is to ensure fairness
between in fast and slow fading environments \cite{Viswanath02}. The
introduction of $\boldsymbol{c}(t)$ changes neither the correlation
property between any pair of columns of $\boldsymbol{B}$ nor the
distribution function of the received SINR.

In what follows, the key point is how to design $\boldsymbol{B}\in
\mathbb{C}^{N_t\times N}$ to accommodate more users (larger $N$) and
maximize system throughput. A straightforward idea is first to
generate a unitary matrix with size $N \times N$, and then choose
its first $N_t$ rows. Despite its simplicity, the main drawback of
this construction is the correlation between different beamforming
vectors $\boldsymbol{b}_{_n}$, $n=1,\,\cdots\,N$ is not guaranteed
at all. Moreover, the transmit power
$\left|\boldsymbol{b}_{_n}\right|^2$, $n=1,\,\cdots,\,N$ at
different beams is randomized.

Now, we present three different methods to construct
$\boldsymbol{B}$ with constrained correlation property.

\subsection{Fourier-Based Construction}
It is well known that a Fourier matrix $\boldsymbol{F} \in
\mathbb{C}^{N_t^2 \times N_t^2}$ is an orthogonal basis in a
$N_t^2$-dimensional complex space. Its projection into a
$N_t$-dimensional complex space forms a tight frame whose elements
have the broadest scattering, and this projection simply retains the
first $N_t$ rows of $\boldsymbol{F}$ \cite{Hochwald00}. Inspired by
this observation, we propose to set the initial beamforming matrix
$\boldsymbol{B}_{_F}$ as, where the subscript $F$ refers to the
Fourier-based beamforming,
\begin{equation} \label{Eq.3-4}
\boldsymbol{B}_{_F}=\frac{1}{\sqrt{N_t}}\begin{bmatrix}
                       1 & 1 & \cdots & 1 \\
                       1 & w & \cdots & w^{N_t^2-1} \\
                       \vdots & \vdots & \ddots & \vdots \\
                       1 & w^{N_t-1} & \cdots & w^{(N_t-1)(N_t^2-1)} \\
                     \end{bmatrix}
\end{equation}
in which $w=e^{-j2\pi/N_t^2}$.

For this choice, the correlation between transmit Beams $l$ and $n$
is
\begin{eqnarray}
c_{_{l,n}} &   = &
\left|{\boldsymbol{B}_{_F}(:\,,\,l)}^H{\boldsymbol{B}_{_F}(:\,,\,n)}\right| \label{Eq.3-5}\\
   &  =  &   \left\{
                       \begin{array}{ll}
                       1, & \hbox{$l=n$} \\
                       \frac{1}{N_t}\left|\frac{\sin{\left(\pi(l-n)/N_t\right)}}{\sin\left(\pi(l-n)/N_t^2\right)}\right|, & \hbox{$l\neq n$}
                       \end{array}
              \right.  \label{Eq.3-5-2}
\end{eqnarray}
Roughly speaking, (\ref{Eq.3-5-2}) suggests the correlation of
$\boldsymbol{B}_{_F}$ behaves like a sinc function and hence all the
cross-correlations between a specific beam and the others are
smaller than its auto-correlation. Therefore, different users'
channels can be well matched by different beamforming vectors.

\begin{figure}
\centering
\includegraphics [width=4in,clip,keepaspectratio]{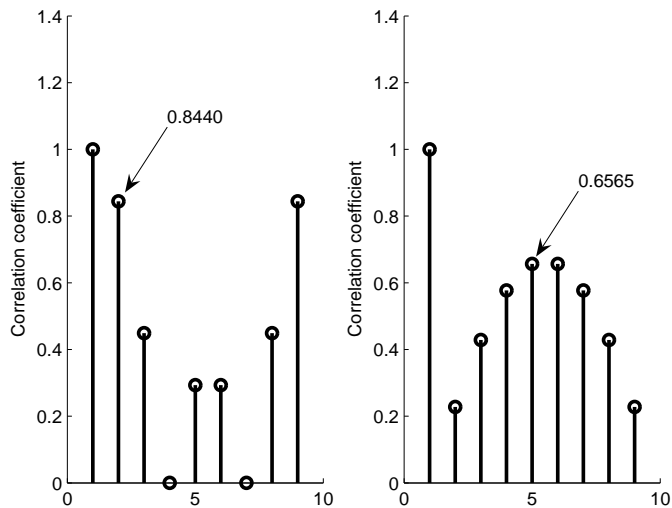}
\caption{The correlation coefficient $c_{_{l,n}}$ shown in
(\ref{Eq.3-5-2}) of Fourier-based constructions with $N_t=3$ and
$N=9$, as a function of $|l-n|$. The left-hand panel corresponds to
the beamforming matrix composed of the first three rows of Fourier
matrix with size $9\times 9$. The right-hand panel refers to our
optimal beamforming matrix with selected $\{3,\,7,\,9\}$ rows.
Obviously, the maximum cross-correlation is decreased from $0.8440$
to $0.6565$.} \label{Fig.2}
\end{figure}

However, it is not necessarily constrained to choose the first $N_t$
rows, but instead the maximum cross-correlation between different
beams can be further lowered by appropriately choosing another set
of $N_t$ components. Unfortunately, the optimal choice with the
lowest maximum cross-correlation
\begin{equation}\label{Eq.3-6}
\delta = \min\max\limits_{l\ne n}{c_{_{l,n}}}
\end{equation}
requires exhaustive searching \cite{Hochwald00}. In Table
\ref{Table1}, we list the number of selected rows with the minimum
$\delta(\boldsymbol{B}_{_F})$ as shown in (\ref{Eq.3-6}), where
$\delta_0$ stands for the maximum cross-correlation with the fist
$N_t$ rows. For example, when $N_t=3$ and $N=9$, $c_{_{l,n}}$ shown
in (\ref{Eq.3-5-2}) is plotted in Fig.~\ref{Fig.2}. We observe that,
with the best choice of $\{3,\,7,\,9\}$ rows, the maximum
cross-correlation is decreased from $0.8440$ to $0.6565$.

\begin{table}[tbp] \label{Table1}
\caption{The comparison of minimum maximum cross-correlation of the
Fourier, Grassmannian, and MUB-based constructions. The
Grassmannian-based one has the best performance if $N=4,\,7,\,13$.
Although only Fourier-based one functions if $N=9$, the MUB-based
one outperforms it if $N=16$.}
\begin{center}
\begin{tabular}{|c|c|c|c|c|c|c|c|c|c|}
 \hline
  $N_t$  & $N$   & $\delta_0$  & \# selected rows   & $\delta(\boldsymbol{B}_{_F})$      &  $\delta(\boldsymbol{B}_{_G})$
  &  $\delta(\boldsymbol{B}_{_M})$   &   Lower bound   & $\hat{\delta}^2$\\
  \hline
  2      & 4     & 0.7071      & \{2, 3\}           & 0.7071   &  \emph{0.5774}        &  0.7071        &  0.5774   &  1    \\
  \hline
  3      & 7     &  0.7490     & \{1, 2, 4\}        & 0.4714   &  \emph{0.4714}        &  $\backslash$  &  0.4714   &  1.3333    \\
  \hline
  3      & 9     &  0.8440     & \{3, 7, 9\}        & \emph{0.6565}   &  $\backslash$  &  $\backslash$  &  0.5      &  2     \\
  \hline
  4      & 13    &  0.8597     & \{1, 3, 4, 8\}     & 0.4330   &  \emph{0.4330}        &  $\backslash$  &  0.4330   &  2.2499     \\
  \hline
  4      & 16    &  0.9061     & \{1, 10, 12, 13\}  & 0.5817   &  $\backslash$  &  \emph{0.5}           &  0.4472   &  3    \\
  \hline
\end{tabular}
\end{center}
\label{Table1}
\end{table}

\subsection{Grassmannian-Based Construction}
Our intention to find $\boldsymbol{B}\in \mathbb{C}^{N_t\times N}$
that has the minimum maximum cross-correlation between any pair of
$N$ beamforming vectors, is equivalent to the Grassmannian line
packing problem in the space $\mathbb{C}^{N_t}$, which is to find a
set of $N$ lines that the minimum distance between any pair of lines
is as large as possible \cite{Strohmer03}. Although the Grassmannian
packing methodology has already been widely applied in the codebook
design \cite{Love03,Mukkavilli03,Love05,Mondal07}, it has seldom
been employed in the design of opportunistic beamforming. To the
best of authors' knowledge, only the tight Grassmannian frames are
exploited to construct the beamforming matrix in the case with
$N=N_t+1$ \cite{Zorba08}. In this subsection, however, we focus on
the generalized cases with $N_t<N \le N_t^2$, making the connections
between Grassmannian frames and opportunistic beamforming design
more transparent.\footnote{Note please that the number of lines $N$
is of no any constraint in the separable infinite-dimensional
Hilbert space for the Grassmannian line packing problem
\cite{Strohmer03}. But in our opportunistic beamforming design, $N
\le N_t^2$ is imposed because of the limited degrees of freedom in
the downlink of MISO systems \cite{Yu06}.}

The Grassmannian frame $\{{\boldsymbol{b}_{_n}}\},\,n=1,\cdots, N$
minimizes the maximum correlation between frame elements among all
unit norm frames which have the same redundancy defined by
\begin{equation}  \label{Eq.3-7}
\eta=\frac{N}{N_t}
\end{equation}
Furthermore, if ${\boldsymbol{b}_{_n}}\in \mathbb{C}^{N_t\times
1},\,n=1,\cdots, N$, then the maximum frame correlation is lower
bounded by \cite[Theorem 2.3]{Strohmer03}
\begin{equation}  \label{Eq.3-8}
\delta=\min\max\limits_{l\ne
n}|\boldsymbol{b}_{_l}^H\boldsymbol{b}_{_n}| \ge
\sqrt{\frac{N-N_t}{N_t(N-1)}}
\end{equation}
Moreover, the equality in (\ref{Eq.3-8}) can only hold if $N\le
N_t^2$, and also now $\{\boldsymbol{b}_{_n}\},\,n=1,\cdots, N$ is an
equiangular tight frame. One achieving the equality in
(\ref{Eq.3-8}) is called optimal Grassmannian frame. Unfortunately,
although there are at most $N_t^2$ degrees of freedom in the
downlink of MISO systems, the optimal Grassmannian frame does not
always exist for any choices of $N_t$ and $N$. For example, if
$N_t=3$, there are at most $N=7$ frame elements for the optimal
Grassmannian frame. In what follows, we give the initial beamforming
matrix with maximum redundancy $\eta$, that is, the number of
transmit beams $N$ is maximized while achieving the equality in
(\ref{Eq.3-8}), according to the following Lemma \ref{Lemma1}.

\newtheorem{lemma}{Lemma}
\begin{lemma} \label{Lemma1}
\emph{(K\"onig \cite{Konig99})} Let $p$ be a prime number and $l$
over the field $\mathbb{N}$ of positive numbers, we set $N_t=p^l+1$
and $N=N_t^2-N_t+1$. Then there exist integers $0\le d_1< \cdots<
d_{N_t}<N$ such that all numbers $1, \cdots, N-1$ occur as residues
mod $N$ of the $N_t(N_t-1)$ differences $d_i-d_q,\,i\ne q,\,1\le
i,\, q \le N_t$. For $n=1, \cdots, N$, we define
\begin{equation}  \label{Eq.3-9}
\boldsymbol{b}_{_n}= \frac{1}{\sqrt{N_t}}\begin{bmatrix}
      e^{j2\pi nd_1/N} & e^{j2\pi nd_2/N} & \cdots & e^{j2\pi nd_{N_t}/N} \\
    \end{bmatrix}^T
\end{equation}
and then the vectors $\boldsymbol{b}_{_n},\,n=1, \cdots, N$ form a
harmonic optimal Grassmannian frame with maximum frame correlation
$\sqrt{N_t-1}/N_t$.
\end{lemma}

\subsubsection{$N_t=2$, $N=4$} In this case, according to \cite[Table
\uppercase\expandafter{\romannumeral2}]{Love03}, the initial
beamforming matrix $\boldsymbol{B}_{_G}$ where the subscript $G$
denotes the Grassmannian-based beamforming, can be given by
\begin{equation} \label{Eq.3-10}
\boldsymbol{B}_{_G}=\left[
                       \begin{array}{cccc}
                         -0.1612-0.7348j & -0.0787-0.3192j & -0.2399+0.5985j &     -0.9541  \\
                         -0.5135-0.4128j & -0.2506+0.9106j & -0.7641-0.0212j &     0.2996   \\
                       \end{array}
                     \right]
\end{equation}

We can easily verify that the columns of $\boldsymbol{B}_{_G}$ form
an equiangular unit norm frame. Furthermore, the equality of lower
bound in (\ref{Eq.3-8}) is attained with
$\delta(\boldsymbol{B}_{_G})=0.5774$. Therefore, the columns of
$\boldsymbol{B}_{_G}$ in (\ref{Eq.3-10}) make an optimal
Grassmannian frame.

\subsubsection{$N_t=3$, $N=7$}
We get $d_1=0$, $d_2=1$, and $d_3=5$ through exhaustive searching,
and then substituting them into (\ref{Eq.3-9}), we have
\begin{equation} \label{Eq.3-11}
\boldsymbol{B}_{_G}=\left[
                       \begin{array}{cccc}
        0.5774    &      0.3600+0.4514j      &     -0.1285-0.5629j      \\
        0.5774    &     -0.1285+0.5629j      &     -0.5202+0.2505j      \\
        0.5774    &     -0.5202+0.2505j      &      0.3600+0.4514j      \\
        0.5774    &     -0.5202-0.2505j      &      0.3600-0.4514j      \\
        0.5774    &     -0.1285-0.5629j      &     -0.5202-0.2505j      \\
        0.5774    &      0.3600-0.4514j      &     -0.1285+0.5629j      \\
        0.5774    &         0.5774           &          0.5774          \\
                        \end{array}
                        \right]^T
\end{equation}
which achieves the equality in (\ref{Eq.3-8}) with
$\delta(\boldsymbol{B}_{_G})=0.4714$.

\subsubsection{$N_t=4$, $N=13$}
We show that $d_1=0$, $d_2=1$, $d_3=3$, and $d_4=9$ through
exhaustive searching, and then substituting them into
(\ref{Eq.3-9}), we get
\begin{equation} \label{Eq.3-12}
\boldsymbol{B}_{_G}=\left[
                       \begin{array}{cccc}
        0.5    &      0.4427 + 0.2324j     &     0.0603 + 0.4964j      &     -0.1773 - 0.4675j      \\
        0.5    &      0.2840 + 0.4115j     &    -0.4855 + 0.1197j      &     -0.3743 + 0.3316j      \\
        0.5    &      0.0603 + 0.4964j     &    -0.1773 - 0.4675j      &      0.4427 + 0.2324j      \\
        0.5    &     -0.1773 + 0.4675j     &     0.4427 - 0.2324j      &      0.0603 - 0.4964j      \\
        0.5    &     -0.3743 + 0.3316j     &     0.2840 + 0.4115j      &     -0.4855 + 0.1197j      \\
        0.5    &     -0.4855 + 0.1197j     &    -0.3743 + 0.3316j      &      0.2840 + 0.4115j      \\
        0.5    &     -0.4855 - 0.1197j     &    -0.3743 - 0.3316j      &      0.2840 - 0.4115j      \\
        0.5    &     -0.3743 - 0.3316j     &    -0.2840 - 0.4115j      &     -0.4855 - 0.1197j      \\
        0.5    &     -0.1773 - 0.4675j     &     0.4427 + 0.2324j      &      0.0603 + 0.4964j      \\
        0.5    &      0.0603 - 0.4964j     &    -0.1773 + 0.4675j      &      0.4427 - 0.2324j      \\
        0.5    &      0.2840 - 0.4115j     &    -0.4855 - 0.1197j      &     -0.3743 - 0.3316j      \\
        0.5    &      0.4427 - 0.2324j     &     0.0603 - 0.4964j      &     -0.1773 + 0.4675j      \\
        0.5    &            0.5            &            0.5            &             0.5            \\
                       \end{array}
                        \right]^T
\end{equation}
which achieves the equality in (\ref{Eq.3-8}) with
$\delta(\boldsymbol{B}_{_G})=0.4330$.

\newtheorem{remark}{Remark}
\begin{remark}  \label{Remark1}
With the Fourier-based construction, we minimize the maximum
cross-correlation between different transmit beams. On the other
hand, different transmit beams are forced to be equiangular with the
Grassmannian-based construction, and also they have the maximum
distance between any pair of beams. However, we claimed in Section
\ref{Section 1} that there are at least $N-N_t$ transmit beams that
are no longer orthogonal with the others. Therefore, a natural
question to ask is: Can we design a beamforming matrix
$\boldsymbol{B}\in C^{N_t \times N}$ with $N_t$ orthogonal vectors
while simultaneously they have the same cross-correlation with the
rest $N-N_t$ vectors? The answer is yes, but we have to rely on the
concept of mutually unbiased bases (MUB) elaborated in the next
subsection.
\end{remark}

\subsection{MUB-Based Construction}
Let $\boldsymbol{U}=\{\boldsymbol{u}_{_1}, \cdots,
\boldsymbol{u}_{_{N_t}}\}$ and
$\boldsymbol{V}=\{\boldsymbol{v}_{_1}, \cdots,
\boldsymbol{v}_{_{N_t}}\}$ be orthonormal bases of
$\mathbb{C}^{N_t}$, $\boldsymbol{U}$ and $\boldsymbol{V}$ are
mutually unbiased if the cross-correlation of vectors satisfies
\begin{equation} \label{Eq.3-3-1}
|\boldsymbol{u}_{_l}^H\boldsymbol{v}_{_n}|=\frac{1}{\sqrt{N_t}},\quad
1\le l,\,n \le N_t
\end{equation}
Furthermore, the set $\mathcal{B}=\{\boldsymbol{U}_1, \cdots,
\boldsymbol{U}_s\}$ is known as an MUB. It is reported in
\cite{Gow07} that $\mathcal{B}$ can be constructed according to the
following Lemma \ref{Lemma2}.

\begin{lemma} \label{Lemma2}
\emph{(Gow \cite{Gow07})} Let $N_t$ be a power of $2$ and let $
\boldsymbol{X}$ consisting of unitary matrices be an irreducible
complex representation of $G_{_{N_t}}$ of degree $N_t$, where
$G_{_{N_t}}$ denotes a finite group of order $N_t^4$. Let $
\boldsymbol{D}$ be a $N_t\times N_t$ matrix that satisfies $
\boldsymbol{D}^{N_t+1}=\boldsymbol{I}$ and $ \boldsymbol{D}^{-1}
\boldsymbol{X}(x)
\boldsymbol{D}=\boldsymbol{X}\left(\delta(x)\right)$ for all $x$ in
$G_{_{N_t}}$. Then the powers $ \boldsymbol{D}$, $\boldsymbol{D}^2$,
$\cdots$, $\boldsymbol{D}^{N_t+1}=\boldsymbol{I}$ define $N_t+1$
pairwise mutually unbiased bases. Furthermore, all entries of $
\boldsymbol{D}$ are in the field $\mathbb{Q}(\sqrt{-1})$.
\end{lemma}

Based on Lemma \ref{Lemma2}, our initial beamforming matrix
$\boldsymbol{B}_{_M}$ in which the subscript $M$ denotes the
MUB-based beamforming, can be given by
\begin{equation}  \label{Eq.3-3-2}
\boldsymbol{B}_{_M}=\left[\begin{array}{cccc}
                            \boldsymbol{D} &  \boldsymbol{D}^2 & \cdots &  \boldsymbol{D}^{N_t}
                          \end{array}
    \right]
\end{equation}

Note that $ \boldsymbol{D}^{N_t+1}=\boldsymbol{I}$ corresponding to
the case of transmit antenna selection is abandoned, due to the
limitation of $N_t^2$ degrees of freedom in the downlink of MISO
systems. Unfortunately, for the cases under consideration with
$N_t\le 4$, the powerful Lemma \ref{Lemma2} can only be exploited in
the cases with $N_t=2,\,4$, rather than the case with $N_t=3$.

\subsubsection{$N_t=2$, $N=4$} In this case, $ \boldsymbol{D}$ is
given by \cite{Gow07}
\begin{equation} \label{Eq.3-3-3}
\boldsymbol{D}=\frac{1+j}{2}\left[
                       \begin{array}{cc}
                         -1 & j   \\
                          1 & j   \\
                       \end{array}
                     \right]
\end{equation}
Substituting it into (\ref{Eq.3-3-2}), we have
\begin{equation} \label{Eq.3-3-4}
\boldsymbol{B}_{_M}=\frac{1+j}{2}\left[
                       \begin{array}{cccc}
                         -1 & j &  j & -j    \\
                          1 & j & -1 & -1    \\
                       \end{array}
                     \right]
\end{equation}

\subsubsection{$N_t=4$, $N=16$} Based on \cite{Heath06}, we can easily show that $\boldsymbol{D}$
can be given by
\begin{equation} \label{Eq.3-3-5}
\boldsymbol{D}=\frac{1}{2}\left[
                       \begin{array}{cccc}
                         -j &  -j  &  -j  &  -j   \\
                          1 &  -1  &   1  &  -1   \\
                         -j &  -j  &   j  &   j   \\
                         -1 &   1  &   1  &  -1   \\
                       \end{array}
                     \right]
\end{equation}
Substituting it into (\ref{Eq.3-3-2}) yields
\begin{equation} \label{Eq.3-3-6}
\boldsymbol{B}_{_M}=\frac{1}{2}\left[
                       \begin{array}{cccccccccccccccc}
                         -j &  -j  &  -j  &  -j  &  -1  &  -1  &  -j  &  j  &  -1  &  j  &  j  &  1  &  j  &  1  &  j  &  -1 \\
                          1 &  -1  &   1  &  -1  &  -j  &  -j  &  -1  &  1  &  -1  &  j  & -j  & -1  &  j  & -1  &  j  &   1 \\
                         -j &  -j  &   j  &   j  &  -j  &   j  &  -1  & -1  &   j  & -1  & -1  & -j  &  j  &  1  & -j  &   1 \\
                          1 &   1  &   1  &  -1  &   1  &  -1  &   j  &  j  &  -j  &  1  & -1  & -j  &  j  & -1  & -j  &  -1 \\
                       \end{array}
                     \right]
\end{equation}

In Table \ref{Table1}, the minimum maximum cross-correlations
$\delta(\boldsymbol{B}_{_G})$ and $\delta(\boldsymbol{B}_{_M})$ of
Grassmannian and MUB-based constructions, respectively, as well as
the lower bound shown in (\ref{Eq.3-8}) are also listed,  with
respect to different number of transmit antennas $N_t$ and number of
transmit beams $N$.

\section{Asymptotic Throughput Analysis} \label{Section 4}

In this section, we investigate the system throughput and thus give
definite answer to which kind of beamforming scheme is most
preferable for a specific $(N_t,\,N)$  configuration, among Fourier,
Grassmannian and MUB-based constructions.

\subsection{Received SINR of User $k$}
For any two non-orthogonal beamforming vectors $\boldsymbol{b}_{_l}$
and $\boldsymbol{b}_{_n}$ where $l\ne n$, $\boldsymbol{b}_{_l}$ can
be expressed in reference to $\boldsymbol{b}_{_n}$ through their
cross-correlation coefficient $\delta_{_l}$, that is,
\begin{equation}\label{Eq.4-2}
\boldsymbol{b}_{_l}=\delta_{_l}
\boldsymbol{b}_{_n}+\sqrt{1-\delta_{_l}^2}\boldsymbol{b}_{_n}^{\bot},\quad
1\le l,\,n \le N
\end{equation}
where $\boldsymbol{b}_{_n}^{\bot}$ stands for the orthonormal vector
to $\boldsymbol{b}_{_n}$. Substituting (\ref{Eq.4-2}) into
(\ref{Eq.2-7}), the received SINR of User $k$ at Beam $n$ can be
rewritten as
\begin{eqnarray}
\gamma_{_{n,\,k}} &   =   &
\frac{\frac{\rho}{N}|\boldsymbol{h}_{_k}\boldsymbol{b}_{_n}|^2}{1+\frac{\rho}{N}\sum\limits_{l=1,\,l\neq
n}^{N}
{|\delta_{_l}\boldsymbol{h}_{_k}\boldsymbol{b}_{_n}+\sqrt{1-\delta_{_l}^2}\boldsymbol{h}_{_k}\boldsymbol{b}_{_n}^{\bot}|^2}}
 \label{Eq.4-3} \\
&   \approx   &
\frac{\frac{\rho}{N}|\boldsymbol{h}_{_k}\boldsymbol{b}_{_n}|^2}{1+\frac{\rho}{N}\hat{\delta}^2
 {|\boldsymbol{h}_{_k}\boldsymbol{b}_{_n}|^2}} \label{Eq.4-4}
\end{eqnarray}
where we explored the approximation
$\boldsymbol{h}_{_k}\boldsymbol{b}_{_n}^{\bot}\approx 0$ in
(\ref{Eq.4-4}), with the assumption that the beamforming vector
$\boldsymbol{b}_{_n}$ matches perfectly with the channel
$\boldsymbol{h}_{_k}$ when the number of active users is large
enough. The parameter $\hat{\delta}^2$ is a constant determined by
the correlation structure of the beamforming matrix, which can be
calculated respectively as follows.
\subsubsection{$N_t=2$, $N=4$} In this case, we observe from Table
\ref{Table1} that the Grassmannian-based beamforming is better than
Fourier or MUB-based construction, since its minimum maximum
cross-correlation $\delta(\boldsymbol{B}_{_G})$ achieves the lower
bound $0.5774$. Furthermore, the Grassmannian-based beamforming
matrix $\boldsymbol{B}_{_G}$ in (\ref{Eq.3-10}) is equiangular, so
that
\begin{equation}
\hat{\delta}^2=3\times 0.5774^2=1  \label{Eq.4-5}
\end{equation}

\subsubsection{$N_t=3$, $N=7$} In this case, it is observed from Table
\ref{Table1} that the beamforming matrix with Grassmannian-based
construction has the same performance as that of the Fourier-based
one with selected $\{1,\,2,\,4\}$ rows. They both achieve the low
bound $0.4714$ and thus
\begin{equation}
\hat{\delta}^2=6\times 0.4714^2=1.3333 \label{Eq.4-6}
\end{equation}

\subsubsection{$N_t=3$, $N=9$}
We find from Table \ref{Table1} that only the Fourier-based
construction functions in this case, though
$\delta(\boldsymbol{B}_{_F})=0.6565$ is larger than the lower bound
$0.5$. From the right-hand panel of Fig.~\ref{Fig.2}, we have
\begin{equation}
\hat{\delta}^2=0.2280^2+0.4285^2+0.5774^2+0.6565^2+0.6565^2+0.5774^2+0.4285^2+0.2280^2=2
\label{Eq.4-7}
\end{equation}

\subsubsection{$N_t=4$, $N=13$}
In this case, it is shown in Table \ref{Table1} that the beamforming
matrix with Grassmannian-based construction has the same performance
as that of the Fourier-based one with selected $\{1,\,3,\,4,\,8\}$
rows and the lower bound $0.4330$ is achieved, hence
\begin{equation}
\hat{\delta}^2=12\times 0.4330^2=2.2499 \label{Eq.4-8}
\end{equation}

\subsubsection{$N_t=4$, $N=16$} In this case, we observe from Table
\ref{Table1} that the MUB-based beamforming matrix outperforms the
Fourier-based one, though they both don't arrive at the lower bound
$0.4472$ but $\delta(\boldsymbol{B}_{_M})=0.5$ of the former is much
closer to it than $\delta(\boldsymbol{B}_{_F})=0.5817$ of the
latter. Therefore,
\begin{equation}
\hat{\delta}^2=12\times 0.5^2=3  \label{Eq.4-9}
\end{equation}

All the above values of $\hat{\delta}^2$ are also listed in the last
column of Table \ref{Table1}.

\subsection{Asymptotic Distribution of $N$ Maximum Received SINRs}
At the base station, we arrange the $K$ received SINRs in Set
$\mathcal{S}_n$ corresponding to Beam $n$ as
$\gamma_{_1},\cdots,\gamma_{_K}$ in an ascending
order,\footnote{There are at most $K$ SINR values in Set
$\mathcal{S}_n$ and meanwhile the other $N-1$ sets are all empty,
which means all users simultaneously have their maximum received
SINRs at Beam $n$.} where $\gamma_{_k}$, $k=1,\cdots,K$ has the same
meaning as $\gamma_{_{n,k}}$ in (\ref{Eq.2-7}) but the beam index
$n$ is ignored here for the sake of notation brevity. Then, we turn
to find the limiting distribution of the $N$ upper extremes of order
statistics $\gamma_{_1},\cdots,\gamma_{_K}$, by applying the
asymptotic theory of extreme order statistics.

It is straightforward to show that
$z=|\boldsymbol{h}_{_k}\boldsymbol{b}_{_n}|^2$ in (\ref{Eq.4-4}) is
of the chi-square distribution with two degrees of freedom, that is,
its PDF can be given by
\begin{equation}\label{Eq.4-10}
f_{_Z}(z)=m\exp(-mz),\quad z\geq 0
\end{equation}
Thus, after some manipulations, the PDF and CDF of $\gamma_{_{k}}$
in (\ref{Eq.4-4}) can be shown respectively as,
\begin{equation}\label{Eq.4-11}
f_{_{\Gamma_k}}(\gamma)=
\frac{mN}{\rho(1-\hat{\delta}^2\gamma)^2}\exp{\left[-\frac{mN\gamma}{\rho(1-\hat{\delta}^2\gamma)}\right]},\quad
\gamma<\frac{1}{\hat{\delta}^2}
\end{equation}
and
\begin{equation}\label{Eq.4-12}
F_{_{\Gamma_k}}(\gamma)=
1-\exp{\left[-\frac{mN\gamma}{\rho(1-\hat{\delta}^2\gamma)}\right]},\quad
\gamma<\frac{1}{\hat{\delta}^2}
\end{equation}

Resorting to the well-known von Mises's sufficient conditions in the
asymptotic theory of extreme order statistics \cite{David03,
Galambos87}, we substitute (\ref{Eq.4-11}) and (\ref{Eq.4-12}) into
the growth function defined by
\begin{equation}  \label{Eq.4-12-2}
g(\gamma)=\frac{1-F_{_{\Gamma_k}}(\gamma)}{f_{_{\Gamma_k}}(\gamma)}
\end{equation}
and then it is straightforward to show the limit of the derivative
of $g(\gamma)$ is, as $\gamma\to 1/\hat{\delta}^2$,
\begin{equation}\label{Eq.4-12-3}
\lim\limits_{\gamma\to
1/\hat{\delta}^2}\frac{\mathrm{d}g(\gamma)}{\mathrm{d}\gamma}=0
\end{equation}
Therefore, $F_{_{\Gamma_k}}(\gamma)$ is in the domain of attraction
of Gumbel-type limiting distribution $H_{_{3,0}}(\gamma)$, where
\cite[p.~296]{David03}
\begin{equation}\label{Eq.4-13}
H_{_{3,0}}(\gamma)=\exp\left(-e^{-\gamma}\right)
\end{equation}
That is, the limiting CDF of the maximum received SINR $\gamma_{_K}$
over $\gamma_{_1},\cdots,\gamma_{_K}$ is
\begin{eqnarray}
\lim\limits_{K \to +\infty}F_{_{\Gamma_K}}(\gamma) & = &
\lim\limits_{K \to +\infty}\left[F_{_{\Gamma_k}}(\gamma)\right]^K \label{Eq.4-14} \\
& = &
\lim\limits_{K \to +\infty}\left[1-\exp{\left(-\frac{mN\gamma}{\rho(1-\hat{\delta}^2\gamma)}\right)}\right]^K \label{Eq.4-14-1} \\
& = &H_{_{3,0}}\left(\frac{\gamma-a}{b}\right)  \label{Eq.4-14-2}
\end{eqnarray}
in which we used (\ref{Eq.4-12}) in (\ref{Eq.4-14-1}). Moreover, the
position parameter $a$ is the solution to \cite[Theroem
2.1.3]{Galambos87}
\begin{equation}\label{Eq.4-14-3}
1-F_{_{\Gamma_k}}(a)=\frac{1}{K}
\end{equation}
Substituting (\ref{Eq.4-12}) into (\ref{Eq.4-14-3}) yields
\begin{equation} \label{Eq.4-14-5}
a=\frac{\rho\ln{K}}{mN+\rho\hat{\delta}^2\ln{K}}
\end{equation}
On the other hand, the scale factor $b$ can be obtained as
\cite[Remark 2.7.1]{Galambos87}
\begin{eqnarray}
b & = & g(a) \label{Eq.4-14-6} \\
& = & \frac{1-F_{_{\Gamma_k}}(a)}{f_{_{\Gamma_k}}(a)}  \label{Eq.4-14-7} \\
& = & \frac{\rho mN}{(mN+\rho\hat{\delta}^2\ln{K})^2}
\label{Eq.4-14-8}
\end{eqnarray}

Moreover, based on \cite[Theorem 2.8.1]{Galambos87}, their
respective limiting CDFs of $N$ upper extremes of
$\gamma_{_1},\cdots,\gamma_{_K}$, that is, $\gamma_{_K}$,
$\gamma_{_{K-1}}$, $\cdots$, $\gamma_{_{K-N+1}}$, can be given by,
as $K\to +\infty$,
\begin{equation}\label{Eq.4-17}
F_{_{\Gamma_{K-n+1}}}(a+b\gamma)=\exp{\left(-e^{-\gamma}\right)\sum\limits_{l=0}^{n-1}{\frac{e^{-l\gamma}}{l!}}},\quad
n=1,\cdots,N
\end{equation}
Finally, it is straightforward to show that their limiting PDFs are,
respectively, as $K\to +\infty$,
\begin{equation}\label{Eq.4-18}
f_{_{\Gamma_{_{K-n+1}}}}(a+b\gamma)=\frac{e^{-n\gamma}}{\Gamma{(n)}}\exp{\left(-e^{-\gamma}\right)},\quad
n=1,\cdots,N
\end{equation}
where $\Gamma{(.)}$ refers to the Gamma function.

\begin{figure}
\centering
\includegraphics [width=4in,clip,keepaspectratio]{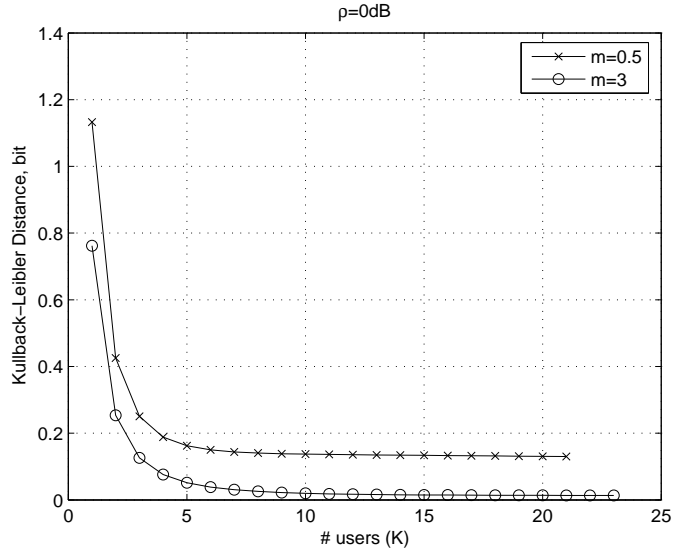}
\caption{The inefficiency of Gumbel-type limiting distribution.}
\label{Fig.3}
\end{figure}

It is well known that the Kullback-Leibler distance
$\mathcal{D}(f||g)$ which behaves like the square of the Euclidean
distance \cite[p.~299]{Cover91} is a measure of the inefficiency of
an approximate distribution $g$ to its true distribution $f$. Thus,
we may exploit it to check the inefficiency of the Gumbel-type
limiting distribution (\ref{Eq.4-14-2}). Specifically, in our
numerical evaluation, we compare the true PDF
\begin{equation}\label{Eq.4-18-2}
f(\gamma)=Kf_{_{\Gamma_k}}(\gamma)\left[F_{_{\Gamma_k}}(\gamma)\right]^{K-1}
\end{equation}
with its limiting PDF
\begin{equation}\label{Eq.4-18-3}
g(\gamma)=\frac{1}{b}\exp{\left(-\frac{\gamma-a}{b}\right)}H_{_{3,0}}\left(\frac{\gamma-a}{b}\right)
\end{equation}
of the maximum received SINR $\gamma_{_K}$, and the Kullback-Leibler
distance is defined as \cite[p.~231]{Cover91}
\begin{equation} \label{Eq.4-19}
\mathcal{D}\left(f||g\right) =\int\limits_0^{+\infty}{f(\gamma)
\log_{_2}{\frac{f(\gamma)}{g(\gamma)}}}
\end{equation}

In Fig.~\ref{Fig.3}, we show the numerical results of
(\ref{Eq.4-19}) with $\rho=0\,\mathrm{dB}$, $m=0.5$ and $3$. We
observe that the Kullback-Leibler distance is only
$0.14\,\mathrm{bits}$ if $K=8$ and $m=0.5$, and it decreases as
increasing $m$.  For example, it is about $0.025\,\mathrm{bits}$ if
$K=8$ and $m=3$. Furthermore, it further decreases as increasing $K$
and finally it approaches zero as $K>23$. Therefore, the Gumbel-type
limiting distribution (\ref{Eq.4-14-2}) is a good approximation to
its true distribution of the maximum received SINR.

\subsection{Throughput Analysis} \label{Section 4-B}
We suppose that all $N$ scheduled users have simultaneously the
maximum SINR at $N$ different transmit beams, then the system
throughput is upper bounded by
\begin{eqnarray}
R_u &  \leq & N\,E\left\{\log_2{(1+\gamma_{_K})}\right\} \label{Eq.4-2-1}\\
& =  &
\frac{N}{b}\int\limits_0^{+\infty}{\log_2{(1+\gamma)}e^{\overline{\gamma}}\exp{\left(-e^{\overline{\gamma}}\right)}}\,\mathrm{d}\gamma
\label{Eq.4-2-2}
\end{eqnarray}
where $\overline{\gamma}=-\left(\gamma-a\right)/b$.

On the other hand, if $N$ scheduled users always have different SINR
at $N$ different transmit beams, that is, if we ignore the small
probability that at least two scheduled users obtain the same SINR,
then the system throughput is lower bounded by
\begin{eqnarray}
R_l & \geq &
E\left\{\sum\limits_{k=K-N+1}^{K}{\log_2{(1+\gamma_{_k})}}\right\}
\label{Eq.4-2-3} \\
& = &
\frac{1}{b}\int\limits_{0}^{+\infty}{\log_2{(1+\gamma)}\left(\sum\limits_{n=1}^{N}\frac{e^{n\overline{\gamma}}}{\Gamma{(n)}}\right)
\exp{\left(-e^{\overline{\gamma}}\right)}}\,\mathrm{d}\gamma
\label{Eq.4-2-4} \\
& = &
\frac{1}{b}\sum\limits_{n=1}^{N}{\frac{1}{\Gamma{(n)}}\int\limits_{0}^{+\infty}{\log_2{(1+\gamma)}e^{n\overline{\gamma}}}
\exp{\left(-e^{\overline{\gamma}}\right)}}\,\mathrm{d}\gamma
\label{Eq.4-2-5}
\end{eqnarray}
in which we used (\ref{Eq.4-18}) in (\ref{Eq.4-2-4}). Unfortunately,
$R_u$ and $R_l$ above can only be calculated by numerical
integration.

Actually, when the number of users is large enough, all $N$
scheduled users have almost the same SINR at $N$ different transmit
beams and hence the system throughput approaches the upper bound.
Therefore, the upper bound $R_u$ shown in (\ref{Eq.4-2-1}) can be
analytically reformulated as
\begin{eqnarray}
R_u  &  \le   & N\,E\left\{\log_2{(1+\gamma_{_K})}\right\} \label{Eq.4-2-6}\\
&  \le  & N\log_2{(1+E\{\gamma_{_K}\})} \label{Eq.4-2-7}\\
& = & N\,\log_2{(1+a+b\Upsilon)} \label{Eq.4-2-8} \\
& = & N\,\log_2{\left[1+\frac{\rho
mN(\Upsilon+\ln{K})+\rho^2\hat{\delta}^2(\ln{K})^2}{(mN+\rho\hat{\delta}^2\ln{K})^2}\right]}
\label{Eq.4-2-9}
\end{eqnarray}
where we used the Jensen's inequality in (\ref{Eq.4-2-7}), and also
in (\ref{Eq.4-2-8}) we explored the Gumbel-type limiting
distribution function as shown in (\ref{Eq.4-14-2}), which has a
mean $\Upsilon=0.5772\cdots$, corresponding to the Euler-Mascheroni
constant. Moreover, we exploited (\ref{Eq.4-14-5}) and
(\ref{Eq.4-14-8}) in (\ref{Eq.4-2-9}).

\begin{remark} \label{Remark2}
In comparison with the orthogonal counterpart with $N=N_t$, we find
from (\ref{Eq.4-3}) that the received SINR in our proposed scheme
with $N_t<N\le N_t^2$ is greatly decreased due to increased
multi-user interferences as well as their mutual non-orthogonality,
which will dramatically deteriorate the system throughput. However,
this deterioration will be compensated by the increased spatial
multiplexing gain $N$, as shown in (\ref{Eq.4-2-9}). Anyway, the
most important characteristic of our proposal is able to serve as
large as $N_t^2$ users simultaneously, which benefits to
significantly decrease the scheduling latency. Moreover, we ignored
the minimum data-rate requirement of each user in this paper. If we
take it into account, the number of simultaneously transmitted users
will possibly be decreased.
\end{remark}

\section{Simulation Results and Discussion} \label{Section 5}
\begin{figure}
\centering
\includegraphics [width=5in,clip,keepaspectratio]{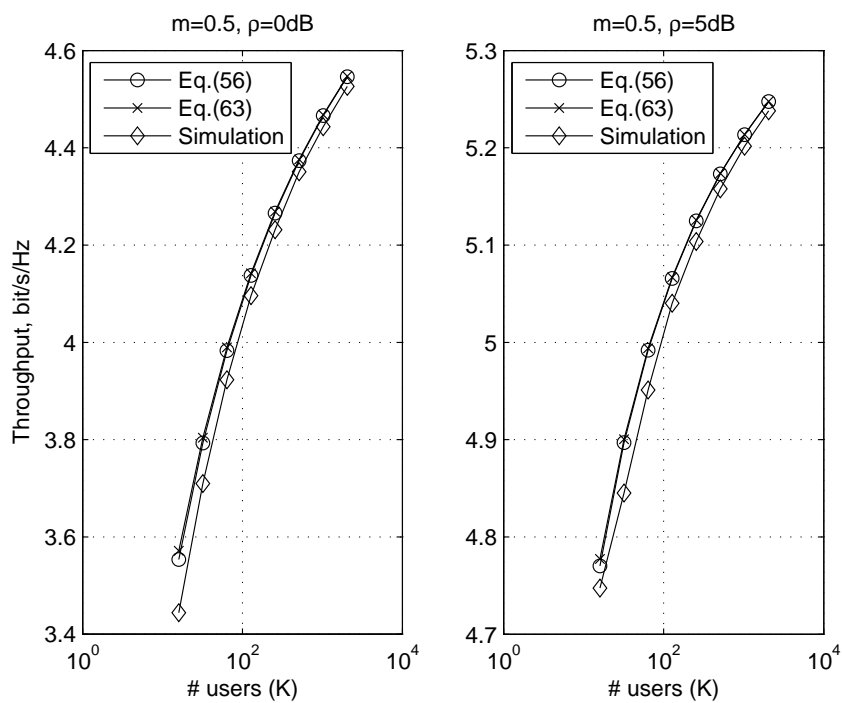}
\caption{The system throughput of OBS with $N_t=3$, $N=7$, and
Grassmannian-based beamforming; $m=0.5$.} \label{Fig.4}
\end{figure}

\begin{figure}
\centering
\includegraphics [width=5in,clip,keepaspectratio]{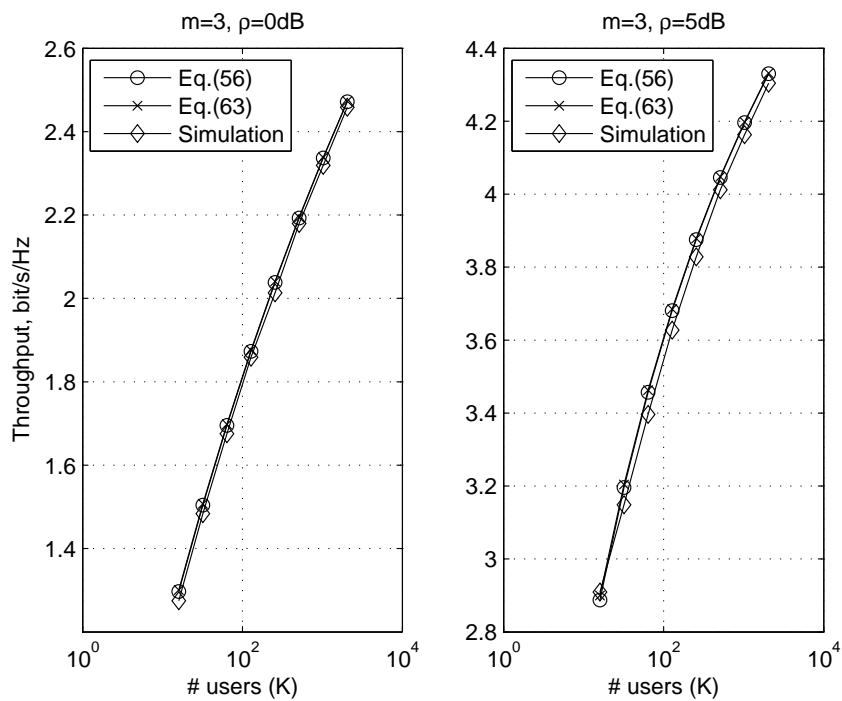}
\caption{The system throughput of OBS with $N_t=4$, $N=16$, and
MUB-based beamforming; $m=3$.} \label{Fig.5}
\end{figure}

\subsection{The Effectiveness of Closed-form Upper Bound}
In this section, we first show the accuracy of our closed-form upper
bound (\ref{Eq.4-2-9}) in comparison with its asymptotic counterpart
(\ref{Eq.4-2-2}) as well as the Monte-Carlo simulation results. In
our simulations, the minimum number of users is $16$, since there
are at most 16 transmit beams if $N_t=4$. On the other hand, the
maximum number of users is set to be $2048$. Although there will not
be so many users in practical cellular communication systems, we are
able to confirm the validity of our throughput analysis by comparing
the numerical results with the simulation ones for such a large
number of users.

In Fig.~\ref{Fig.4}, we show the system throughput of OBS with
Grassmannian-based beamforming matrix as shown in (\ref{Eq.3-11}),
where $N_t=3$, $N=7$, and $m=0.5$. We observe that the closed-form
upper bound (\ref{Eq.4-2-9}) almost always overlaps with the
asymptotic (\ref{Eq.4-2-2}), no matter how many users there are or
whatever SNR is $0$ or $5\,\mathrm{dB}$. However, although we
claimed in Section \ref{Section 4-B} that all the scheduled users
have almost the same maximum SINR as the number of users approaches
infinity, there is always a very small gap between the simulation
results and the upper bound as shown in Figs.~\ref{Fig.4} and
\ref{Fig.5}. For example, when $m=0.5$, $\rho=0\,\mathrm{dB}$ and
the number of users $K=64$, it is observed from the left-hand panel
of Fig.~\ref{Fig.4} that the difference between (\ref{Eq.4-2-9}) and
the simulations results is about $0.06\,\mathrm{bit/s/Hz}$, or
$1.6\%$ in relative to the simulation result
$3.93\,\mathrm{bit/s/Hz}$. Moreover, this gap becomes smaller and
smaller as the number of users or the average SNR increases. The
same observation can be attained from Fig.~\ref{Fig.5}, where the
beamforming matrix is based on MUB construction as shown in
(\ref{Eq.3-3-6}), $N_t=4$, $N=16$, and $m=3$. Therefore, we conclude
that our upper bound (\ref{Eq.4-2-9}) is very tight with simulation
results, and thus it can be exploited below to evaluate the system
throughput effectively.

\subsection{System Throughput of Proposed Schemes}

\begin{figure}
\centering
\includegraphics [width=5in,clip,keepaspectratio]{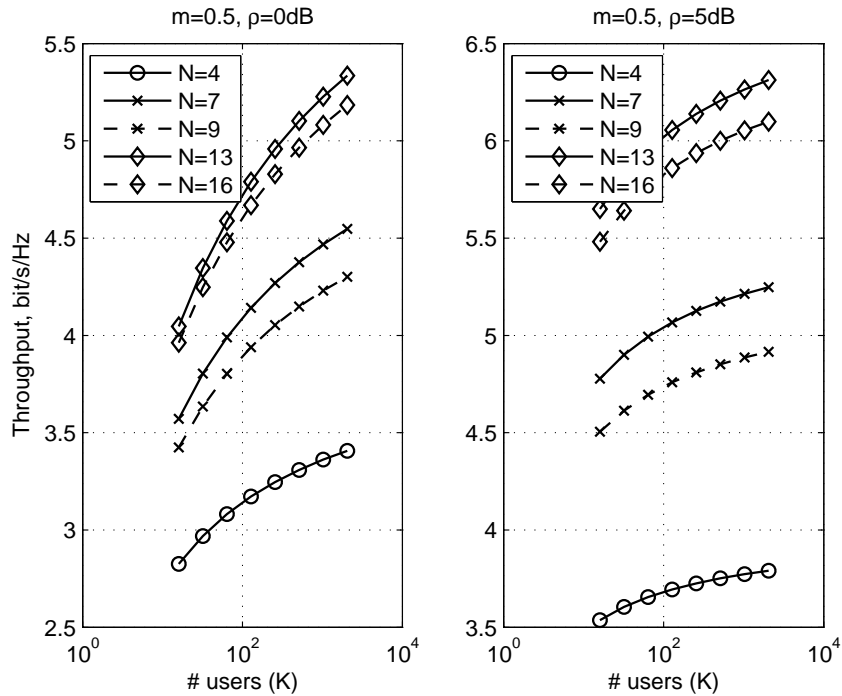}
\caption{The system throughput of OBS with $N=4$, $7$, $9$, $13$,
$16$, and $m=0.5$.} \label{Fig.6}
\end{figure}

\begin{figure}
\centering
\includegraphics [width=5in,clip,keepaspectratio]{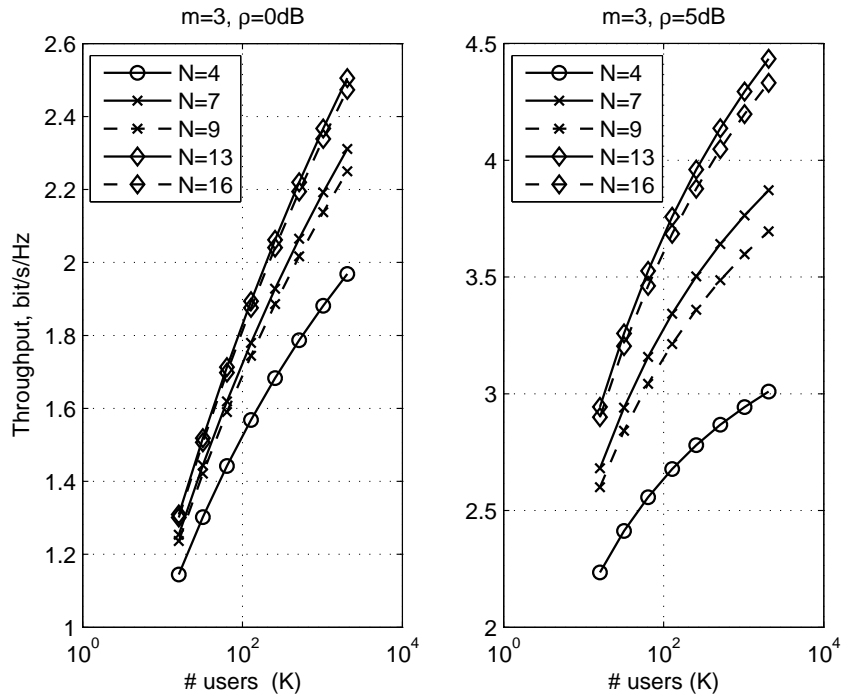}
\caption{The system throughput of OBS with $N=4$, $7$, $9$, $13$,
$16$, and $m=3$.} \label{Fig.7}
\end{figure}

According to (\ref{Eq.4-2-9}), we compare the system throughput of
proposed schemes in Figs.~\ref{Fig.6} and \ref{Fig.7}, where the
beamforming construction of $N=4,\,7,\, 13$ are Grassmannian based,
$N=9$ is Fourier based, and $N=16$ is MUB based, respectively. We
find that the OBS with Grassmannian-based beamforming achieves the
maximum system throughput whenever $N_t=3$ or $4$, corresponding to
$N=7$ or $13$, respectively. But if we want to fully exploit the
degrees of freedom when $N_t=3,\,4$, then we have to rely on the
Fourier and MUB-based construction, that is, $N=9,\,16$ users can be
simultaneously transmitted, respectively. Unfortunately, the
increase of the number of simultaneously scheduled users is at the
penalty of system throughput. For example, when $K=64$ and
$\rho=0\,\mathrm{dB}$, we observe from the left-hand panel of
Fig.~\ref{Fig.6} that the throughput difference between the cases
with $N=7$ and $N=9$ is about $0.19\,\mathrm{bit/s/Hz}$, or $4.8\%$
throughput loss of the case with $N=9$ in relative to the throughput
$3.99\,\mathrm{bit/s/Hz}$ of the case with $N=7$. Moreover, this
throughput loss will slightly increase as the number of users or the
average SNR increases, but it decreases fast as the the variance
$1/m$ of Rayleigh fading decreases by comparing Fig.~\ref{Fig.6}
with Fig.~\ref{Fig.7}. Furthermore, we observe that the system
throughput degrades with decreasing variance $1/m$ by comparing
Fig.~\ref{Fig.6} with Fig.~\ref{Fig.7}. This degradation should not
come as a surprise since the multi-user diversity gain will be
smaller and smaller as the channel fading becomes more and more
stable \cite{Hochwald04}.

\subsection{The Preferred Low SNR Case}

\begin{figure}
\centering
\includegraphics [width=5in,clip,keepaspectratio]{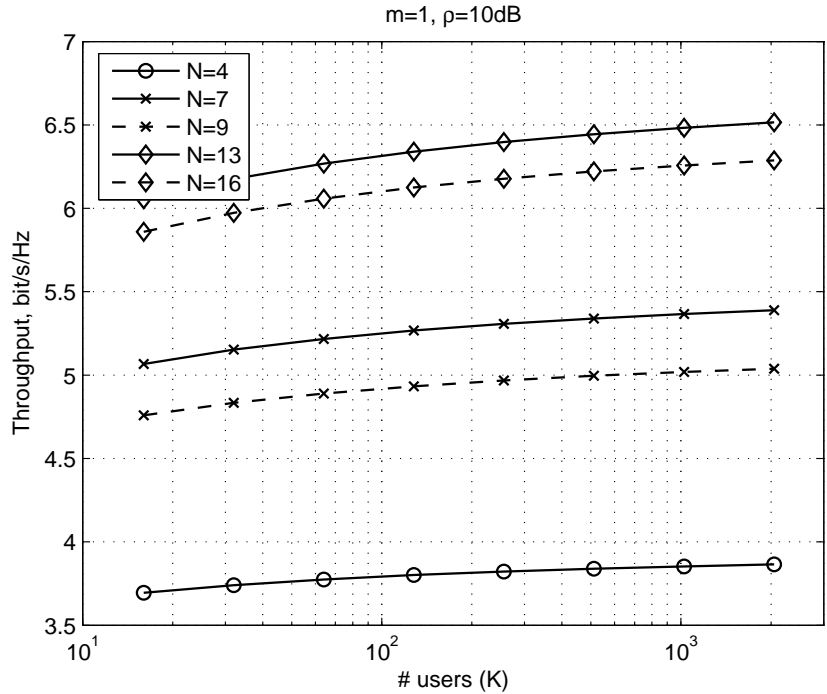}
\caption{The system throughput of OBS with $N=4$, $7$, $9$, $13$,
$16$, $m=1$, and $\mathrm{SNR}=10\,\mathrm{dB}$.} \label{Fig.8}
\end{figure}

We point out that the proposed schemes with $N_t<N\le N_t^2$
transmit beams is much beneficial to the low SNR case over the high
SNR scenario. In Fig.~\ref{Fig.8}, we see that the system throughput
increases very slowly as the number of users increases sharply from
$16$ to $2048$, where $m=1$ and $\rho=10\,\mathrm{dB}$. That is,
when the SNR is high enough, the multi-user diversity gain becomes
saturated soon. This phenomena can be understood as follows: We see
from (\ref{Eq.4-4}) that the received SINR $\gamma_{_{n,\,k}}$ can
be approximated to $1/\hat{\delta}^2$ if $\rho$ is large enough,
that is, the value of $\gamma_{_{n,\,k}}$ is independent of the user
index $k$ and therefore the multi-user diversity gain vanishes.

\subsection{Throughput Comparison With Orthogonal Counterpart}
\begin{figure}
\centering
\includegraphics [width=5in,clip,keepaspectratio]{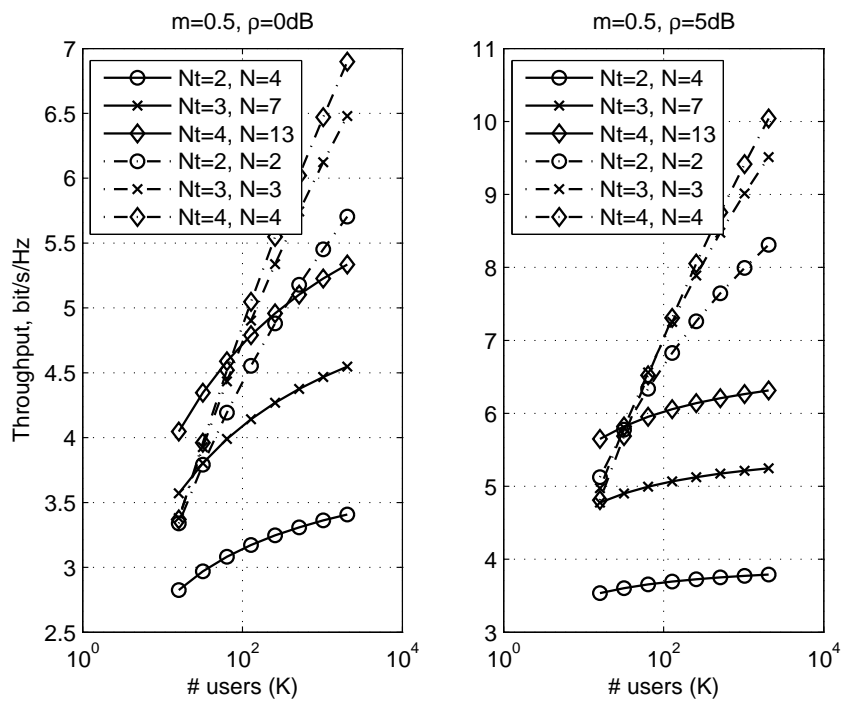}
\caption{Throughput comparison between proposed scheme and their
orthogonal counterpart, $m=0.5$.} \label{Fig.9}
\end{figure}

\begin{figure}
\centering
\includegraphics [width=5in,clip,keepaspectratio]{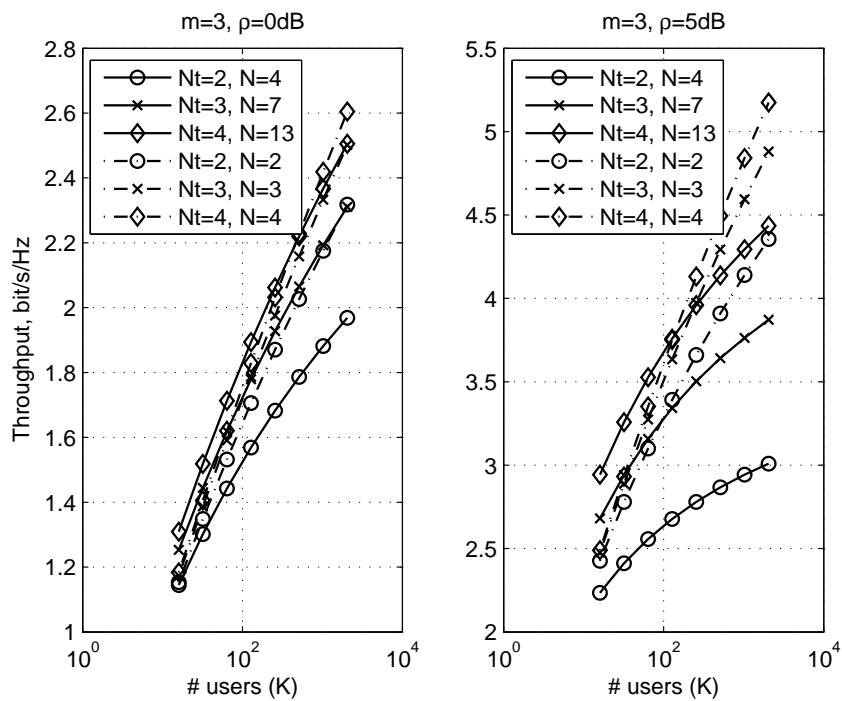}
\caption{Throughput comparison between proposed scheme and their
orthogonal counterpart, $m=3$.} \label{Fig.10}
\end{figure}

Although our proposed schemes can schedule much more users than the
number of transmit antennas $N_t$, how about the system throughput
in comparison with their conventional orthogonal counterparts in
which the number of transmit beams $N$ equals $N_t$? In
Figs.~\ref{Fig.9} and \ref{Fig.10}, we show their system throughput
comparison. Usually, in each cell of a practical cellular
communication system, there are only tens of simultaneously active
users. In this regard, we can see from the right-hand panel of
Fig.~\ref{Fig.9} that the system throughput of our proposed scheme
with $N_t=4,\,N=13$ and its orthogonal counterpart with
$N_t=4,\,N=4$ are $6.06$ and $7.31\mathrm{bit/s/Hz}$, respectively,
if $m=0.5$ and $K=128$. In other words, there is $18\%$ throughput
loss but the scheduled users is of $225\%$ increase! Furthermore,
when the channel becomes more and more flat (as $m$ increases), the
throughput loss turns to be smaller and smaller, and even if $m=3$
and $K\le 128$ as shown in Fig.~\ref{Fig.10}, the system throughput
of our proposed scheme with $N_t=4,\,N=13$ outperforms that of its
orthogonal counterpart as well as any other cases with $N_t<4$. The
underlying reason is that, as the number of active users is small,
for example, in any practical cellular system, the multi-user
diversity gain is strictly limited and thus larger spatial
multiplexing gain of our scheme leads to larger system throughput.
Therefore, the proposed schemes, especially, the case with
$N_t=4,\,N=13$, is of great interest in practical employment.

\begin{remark}  \label{Remark3}
Actually, the proposed schemes can be generalized to the cases with
the number of receive antennas $N_r>1$. Although he/she has the
potential to use up to $N_r^2$ degrees of freedom, we can employ a
combining strategy to reduce effectively each user with $N_r>1$ to a
single-dimensional receive terminal \cite{Yu06,Rhee04}. That is, the
rank of received signal of each user is forced to be $1$, and thus
the number of simultaneously transmitted users remains $N_t^2$ all
the same.
\end{remark}

\section{Conclusion} \label{Section 6}
\par
Inspired by the $N_t^2$ degrees of the freedom in the downlink of
MISO systems, we demonstrated how to transmit to more than $N_t$
users simultaneously, whereas at most $N_t$ users can be
simultaneously scheduled in the conventional MISO beamforming
systems. We proposed three different opportunistic beamforming
schemes: Fourier, Grassmannian, and MUB-based constructions. The
Grassmannian-based scheme achieves the maximum system throughput
with the number of transmit beams $N=4$, $7$, $13$ in the cases with
$N_t=2$, $3$, $4$, respectively, by taking the optimal Grassmannian
frames as the beamforming matrices. However, it can not exploit all
$N_t^2$ degrees of freedom when $N_t>2$. On the other hand, if we
want to fully exploit $9$ and $16$ degrees of freedom in the cases
with $N_t=3$ and $4$, we may resort to the Fourier and MUB-based
schemes, respectively, despite a little penalty on system
throughput. Finally, the special Grassmannian-based case with
$N_t=4$ and $N=13$ was shown to be promising for practical
employment in cellular systems, since it outperforms its orthogonal
counterpart in terms of the number of simultaneously scheduled users
but without any throughput loss.

\vfill
\end{document}